# Amplification of seismic ground motion in the Tunis basin: Numerical BEM simulations vs experimental evidences

M. Kham[1,2], J-F Semblat[1], N. Bouden-Romdhane[3]

[1] EDF R&D, Clamart, France (marc.kham@edf.fr)
[2] Université Paris-Est, IFSTTAR, Dept of Geotechnical Eng., Environment and Risks, Paris, France, (jean-francois.semblat@ifsttar.fr)
[3] ENIT, Tunis, Tunisia

**ABSTRACT**
This paper aims at the analysis of seismic wave amplification in a deep alluvial basin in the city of Tunis in Tunisia. This sedimentary basin is *3000m* wide and *350m* deep. Since the seismic hazard is significant in this area, the depth of the basin and the strong impedance ratio raise the need for an accurate estimation of seismic motion amplification. Various experimental investigations were performed in previous studies to characterize site effects. The *Boundary Element Method* is considered herein to assess the parameter sensitivity of the amplification process and analyse the prevailing phenomena. The various frequencies of maximum amplification are correctly estimated by the BEM simulations. The maximum amplification level observed in the field is also well retrieved by the numerical simulations but, due to the sensitivity of the location of maximum amplification in space, the overall maximum amplification has to be considered. The influence of the wave-field incidence and material damping is also discussed.

## 1 Introduction

Tunis and its area are the location of various seismic events (Bouden-Romdhane & Mechler, 2002). According to the historical seismicity of the site, large and destructive earthquakes are estimated to occur on an average every four centuries, the next being expected at the beginning of *21$^{st}$* century. An explanation for the sensitivity of the site is due to its geotechnical configuration. Tunis city is actually located in an alluvial valley with soft soil layers over a rigid basement, configuration which is known to increase the amplification of seismic ground motion i.e. site effects (Bard and Bouchon, 1985; Luzon et al, 2004; Pitilakis et al, 1999; Semblat et al, 2005).
In order to build the seismic hazard map of the region, two experimental surveys have been conducted, firstly by microtremor recordings (*H/V* Spectral Ratios) using a dense network (*260* measurement points) over the city centre, secondly by seismic events recordings (Standard Spectral Ratios) performed in *7* stations in the area (Bouden-Romdhane & Mechler, 2002). Ten seismic events were recorded: 2 weak local events, 7 teleseismic events and 1 quarry shot. All these experimental data are convergent to give the level and the frequency range of major ground motion amplification expected in the different locations of the area and highlight the key role of such site effects.

These surveys also give interesting validation for theoretical investigations of the phenomena. Thereafter, we propose a numerical model (*Boundary Element Method*) to assess seismic ground motion amplification in Tunis. The analysis of site effects through numerical approaches is interesting since it allows quantitative assessment of both the amplification level and location of its largest values. Furthermore, such models also allow to analyse the influence of other parameters such as frequency, damping, incidence, wavetype, etc.

However, only weak motions were recorded and modelled in this work. In case of strong seismic events, the present investigations should be generalized by considering modulus degradation in the soil, frequency drop in the response and larger energy dissipation during the seismic motion (Bonilla, 2000; Delépine et al, 2009; Field et al, 1997; Hartzell et al, 2004; Idriss, 1990, Santisi d'Avila et al, 2012).

In this paper, after a brief description of the alluvial deposit in Tunis, the main experimental data are recalled. In section 3, the principles of the BEM are summarized. Section 4 presents the BEM model and section 5 compares the computed solutions to experimental data. A preliminary investigation of the role of several parameters such as frequency, incidence and damping is also given.

## 2  Tunis site and experimental data

### 2.1  Alluvial basin

Tunis city spreads in an alluvial basin or valley oriented *WNE-ESE* and surrounded by three geological units (Bouden-Romdhane & Mechler, 2002): in the south, the Sidi Bel Hassen (*88m* NGT) and Jebel Jeloud-Kharrouba (*108m* GNT) hills are oriented *NE-SW*; the west side is settled by a mountain chain oriented NW-SE as shown in Figure **1**; the relief is completed by the north hills oriented *N-S*. From the geological point of view, the basin is composed of an alluvial deposit over a stiffer basement made of limestone. The alluvial filling can also be divided into two types: below, the continental and marine sedimentary deposits of the ancient Quaternary constitute the thickest part of the layer and can be found up to a *360* meters depth; then, the softer surficial part (*60* meters thick) of the layer is made of mud deposits from different rivers during the recent Quaternary. Because of the seismic hazard potential due to the presence of many faults over the region (Figure **2**) and the location of the city over the soft surficial deposit, the seismic hazard in Tunis is significantly increased due to the amplification of the ground motion. It justifies the two experimental surveys conducted recently to assess the seismic risk and to propose a more accurate microzonation of the region.

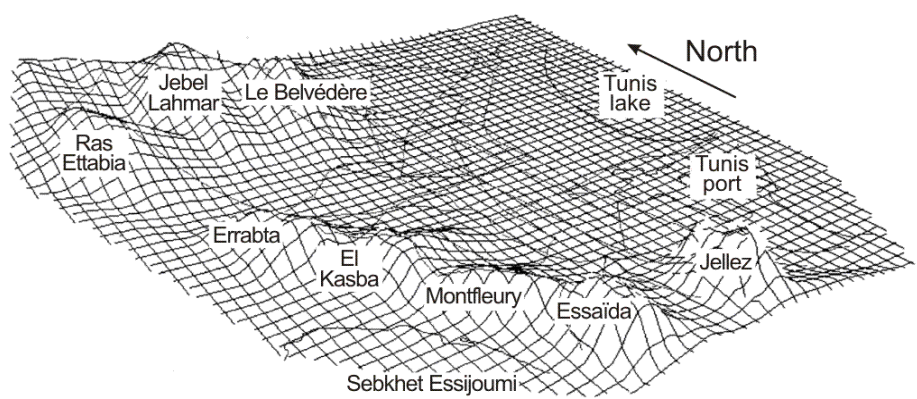

Figure **1**: The major geomorphologic units of Tunis city

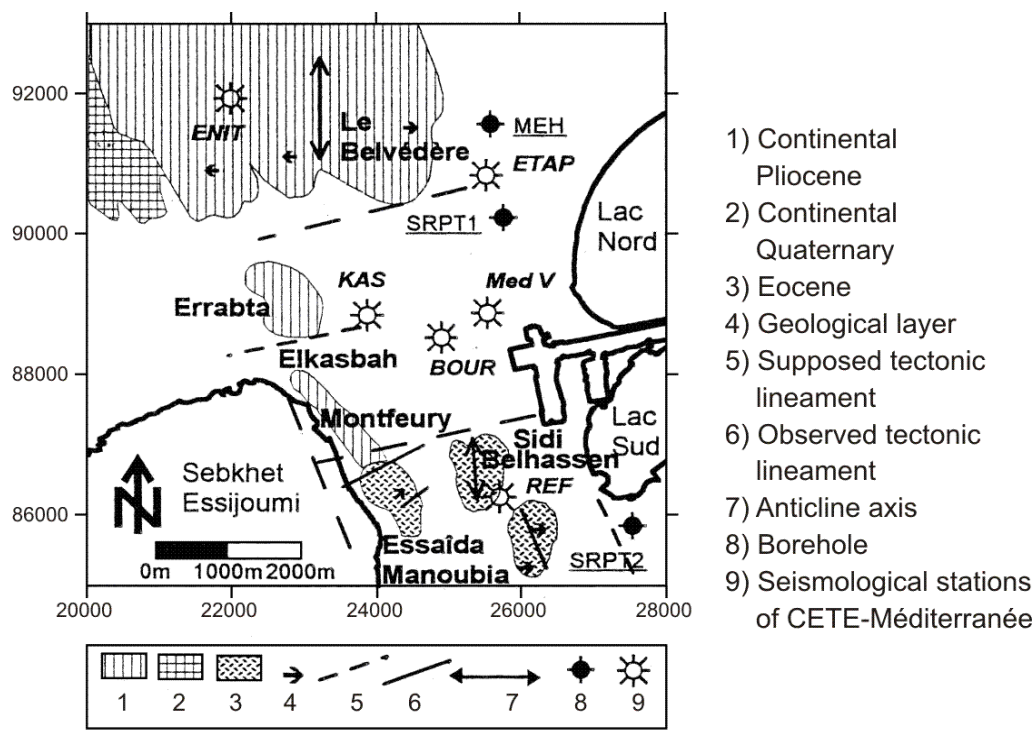

Figure 2: Simplified structural map of Tunis site
and location of the seismological stations.

*2.2 Experimental survey*
A first experimental campaign was led by CETE-Méditerranée and ENIT in 1997 (Bouden-Romdhane & Mechler, 1998) applying the *HVSR* method (microtremor recordings). The method consists in computing the horizontal-to-vertical spectral ratio of surface motion from noise recordings in order to assess the first resonance frequency of the site. This technique has proved to be efficient through previous experiments (Duval, 1994). Furthermore, since the technique is not linked to any seismic event and also easy to manage, a very dense coverage up to *260* measurement points over the city was allowed.
The second campaign is founded on the classical Site to Reference Spectral Ratio method where the transfer function of the site is obtained, from real earthquake recordings, by dividing the spectrum of a local site by a reference spectrum corresponding to the bedrock. These spectral ratios are generally called Standard Spectral Ratios and will be denoted SSR in the following. This technique is well recognised to be efficient for assessing site-effect but is also more difficult to apply than *HVSR*. On the one hand, it is necessary to record a significantly large seismic event and on the other hand a reference bedrock station has to be found. However, measurements were made in Tunis using the *7* stations of the seismological mobile network of CETE-Méditerranée (France). The sensors are located as shown in Figure **2** and named in brief as: *BOUR, Med V, ETAP, LAC, KASB, ENIT* & *REF* (for the reference station).

As shown in Figure **3** (also see (Bouden-Romdhane & Mechler, 2002) for details), both experimental results are converging especially for low frequencies where both the frequency peaks and magnitude of the signals are in good agreement (for higher frequencies, the HVSR method underestimates the amplification magnitude, as expected). They logically reveal higher amplification of the surface motion at locations where the alluvial layer is thicker. These results are very satisfactory since they are in agreement with theoretical expectations. They also constitute a very rich data set for further theoretical or numerical investigations.

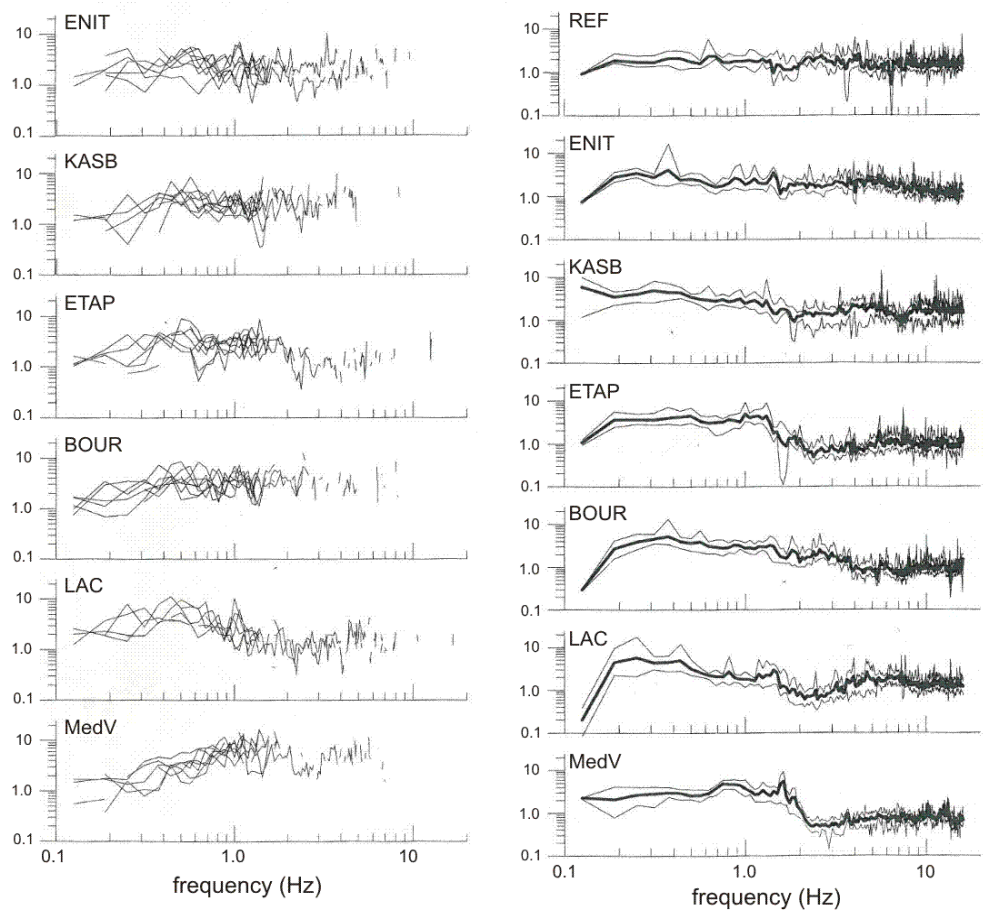

Figure **3**: Standard Spectral Ratios from earthquake recordings (left) and H/V Spectral Ratios from microtremor recordings (right) in the city of Tunis.

## 3   The Boundary Element Method

### 3.1   Numerical methods for wave propagation

To analyze wave propagation (seismic waves, vibrations, etc) in 2D or 3D geological structures, various numerical methods are available:

- the finite difference method is accurate in elastodynamics but is mainly adapted to simple geometries (Moczo et al, 2002; Virieux, 1986),
- the finite element method is efficient to deal with complex geometries and numerous heterogeneities (even for inelastic constitutive models (Bonilla, 2000)) but has several drawbacks such as numerical dispersion and numerical damping (Hughes et al, 2008; Ihlenburg and Babuška, 1995; Semblat et al, 2000a) and (consequently) numerical cost in 3D elastodynamics,
- the spectral element method has been increasingly considered to analyse 2D/3D wave propagation in linear media with a good accuracy due to its spectral convergence properties (Chaljub et al, 2007; Faccioli et al, 1996; Komatitsch and Vilotte, 1998),
- the boundary element method allows a very good description of the radiation conditions but is preferably dedicated to weak heterogeneities and linear constitutive models (Beskos, 1997; Bonnet, 1999; Dangla et al, 2005; Semblat et al, 2000b, 2005). Recent developments have been proposed to reduce the computational cost of the method especially in the high frequency range (Chaillat et al, 2008, 2009; Fujiwara, 2000),
- the scaled boundary finite element method which is a kind of solution-less boundary element method (Wolf 2003).

Furthermore, when dealing with wave propagation in unbounded domains, many of these numerical methods raise the need for absorbing boundary conditions to avoid spurious reflections. Various absorbing layer methods, such as Perfectly Matched Layers-PMLs (Festa and Nielsen, 2003) or Caughey Absorbing Layer Method-CALM (Semblat et al, 2011), have been proposed in the recent years to reduce spurious reflected waves. It is also possible to couple FEM and BEM (Aochi et al, 2005; Bonnet, 1999; Grasso et al, 2012) allowing an accurate description of the near field (FEM model including complex geometries, numerous heterogeneities and nonlinear constitutive laws) and a reliable estimation of the far-field (BEM involving accurate radiation conditions).

## 3.2 Principles of the BEM

The Boundary Element Method is well adapted to analyse wave propagation in infinite or semi-infinite media, because it reduces the whole domain to its boundaries and the computational cost, there is no cumulative numerical dispersion error (i.e. numerical dispersion, as for Finite Difference or Finite Element Methods) and no spurious reflections at the model boundaries. However, the main difficulty of the method is the regularisation of the singular integrals raised in the integral formulation. It may be very uneasy to solve in some cases. The BEM remains nevertheless very interesting in the field of earthquake engineering where most of the problem configurations are unbounded (Dangla et al, 2005).

Let us consider a dynamic stationary problem where all the values are of the same type than the body force $f(x,t)=f(x).exp(-i\omega t)$. The equilibrium equations for the elastic, homogeneous and isotropic medium $D$ of surface $S$ are :

$$div(\underline{\underline{\sigma}}(\underline{x})) + \rho\omega^2 \underline{u}(\underline{x}) + \rho\underline{f}(\underline{x}) = \underline{0} \qquad (1)$$

Let $(\underline{u},\underline{\underline{\sigma}})$ and $(\underline{u}',\underline{\underline{\sigma}}')$ two elastodynamic states verifying Eq (1), the following reciprocity equation can be deduced :

$$\int_S \underline{t}^{(n)}(\underline{x}).\underline{u}'(\underline{x})ds(\underline{x}) + \int_D \rho\underline{f}(\underline{x}).\underline{u}'(\underline{x})dv(\underline{x}) = \int_S \underline{t}'^{(n)}(\underline{x}).\underline{u}(\underline{x})ds(\underline{x}) + \int_D \rho\underline{f}'(\underline{x}).\underline{u}(\underline{x})dv(\underline{x}) \qquad (2)$$

where $\underline{t}^{(n)} = \underline{\underline{\sigma}}.\underline{n}$ with $\underline{n}$ being the normal surface vector.

## 3.3 Integral representation for SH waves :

Let $(\underline{u}',\underline{\underline{\sigma}}')$ be the solution of Eq (1) for the unit body force $\rho\underline{f}'(\underline{x}).\underline{e}_Z = \delta(\underline{x}-\underline{y}).\underline{e}_Z$ and verifying the free surface boundary conditions. These solutions are called the half-space Green functions and are noted $U_Z^\omega(\underline{x},\underline{y})$ and $T_Z^{(n)\omega}(\underline{x},\underline{y})$. They correspond to the antiplane solution at point $\underline{x}$ due to an antiplane unit force concentrated at point $\underline{y}$. Then Eq (2) expresses the integral representation of the SH-solution at any point $\underline{y}$ out of the boundary unknown quantities by a collocation method.

$$I_D(\underline{y})w(\underline{y}) = -\int_S U_Z^\omega(\underline{x},\underline{y}).t_Z^{(n)}(\underline{x})ds(\underline{x}) \qquad (3)$$

where $I_D(\underline{y})$ is 1 when $\underline{y} \in D$ and 0 when $\underline{y} \notin D$.

*3.4 Boundary integral equation :*

Since the half-space Green Functions are singular at point $\underline{y}$, Eq (3) is not defined for $\underline{y} \in S$. It is then necessary to give a regularized expression of the integral representation (3) at points $\underline{y} \in S$. It becomes the following boundary integral equation (Bonnet, 1999, Dangla et al, 2005):

$$C(\underline{y})w(\underline{y}) = -\oint_S U_Z^\omega(\underline{x},\underline{y}).t_Z^{(n)}(\underline{x})ds(\underline{x}) \qquad (4)$$

where $C(\underline{y}) = \lim_{e \to 0} \int_{\gamma_e} T_Z^{(n)\omega}(\underline{x},\underline{y})ds(\underline{x})$ is ½ when $\underline{y}$ belongs to a regular line interface and the symbol $\oint$ means the Cauchy principal value integral.

## 4   Numerical analysis of ground motion amplification

*4.1 Numerical model of the site*

A 2D geological profile of the Tunis basin is chosen. It corresponds to the N-S longitudinal direction of the valley crossing BOUR and REF stations. As a matter of fact, some experimental evidences are arguing for a "2D behaviour" of the basin in this area. Furthermore, the large amount of data available for this cross section allows us to compare the numerical results to both H/V and SSR methods, especially for BOUR station as shown in Figure 4. The 3D knowledge of the surface geology is rather poor.

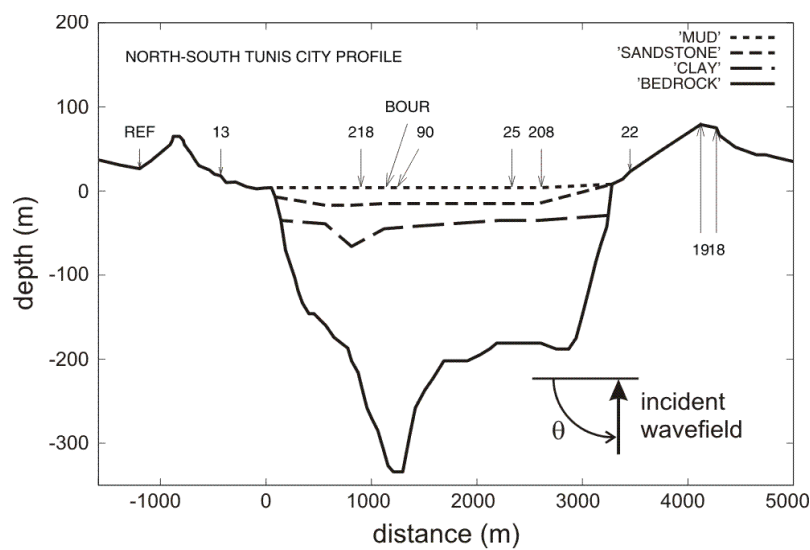

Figure 4: North-South profile used for the model: mud: $V_S$=60;100;150m/sec - d=1.6; sandstone: 400<$V_S$<700m/sec - d=1.85; clay: $V_S$=500;600;650m/sec - d=1.8; bedrock: $V_S$=1400m/sec – d=2.2

The geotechnical properties of the different surficial layers were derived from the prospective borehole at site SRPT1 (Bouden-Romdhane and Mechler, 2002). Hence, in our model, we first consider the following simplified assumptions:

1) The surficial layer is homogeneous with $V_S$=500m/sec, d=1.8 and the properties of the elastic bedrock are $V_S$=1400m/sec, d=2.2.
2) The medium is undamped, which means Q=0.
3) The excitation is considered as an antiplane shear plane wave (SH), which may be physically translated as a E-W polarized ground motion.

The BEM simulations are performed in the frequency domain (time-harmonic problem) and the amplification factor at the free surface is estimated (free surface effect being removed).

### 4.2 Amplification estimated versus distance and frequency

Since the aim of the experimental surveys is to determine areas of large amplification within the region, it is interesting to investigate the distribution of motion amplification along the free surface of the basin with respect to frequency. As suggested by the work of Semblat et al (2005) in the case of Volvi (Greece), Figure 5 reveals that the resonance frequencies of the layer are linked to the local thickness of the deposit: the resonance frequency of the thickest part is about 0.4Hz where the ground motion amplification can reach nearly 7 while the thinnest parts on both sides have resonance frequencies ranging between 0.5 and 1Hz with a maximum amplification of 7.5 at 0.6Hz. For higher frequencies, it is worth to notice that, for specific frequency ranges, we obtain periodically spaced amplification peaks all along the free surface. Conversely, for other frequencies such as 1.1Hz, 1.4Hz and above 2.25Hz it seems that there is no significant site effect on the free surface. One explanation is the alternative constructive and destructive wave interferences within the irregular geometry of the layers. This points out the interest of the numerical approach since it is then possible to have access to qualitative and quantitative information versus specific parameters and associated to other complex phenomena difficult to investigate in the field.

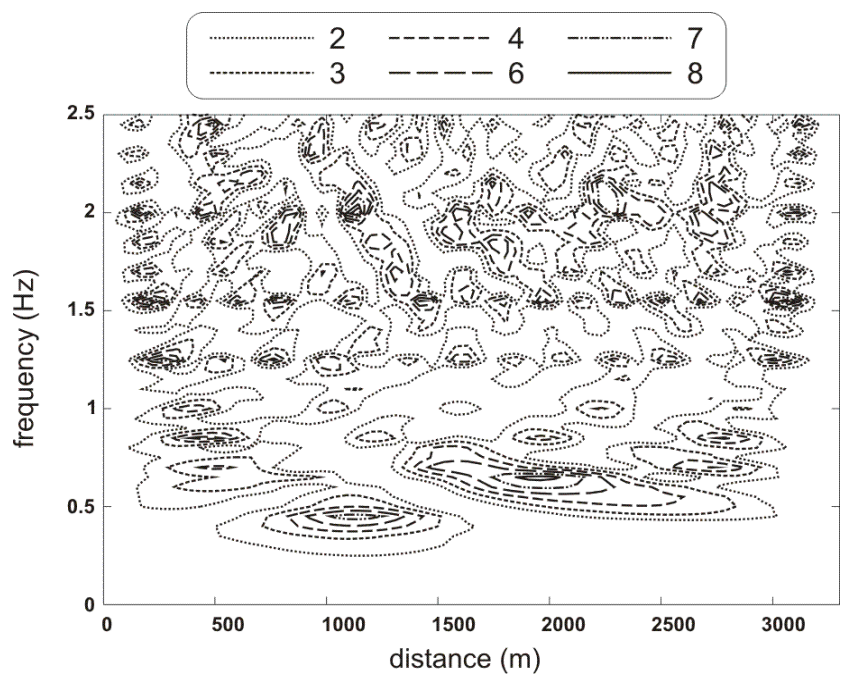

Figure 5: Amplification factor versus frequency and distance along the surface.

*4.3   Influence of the wavefield incidence*

Another interesting point is the influence of the wave-field incidence on the amplification factor along the free surface. It gives an insight on the influence of the seismic source location on site effects. The variation of the amplification factor along the surface are displayed for various incidences angles (50° to 110° by steps of 10°) at f=0.4Hz in Figure 6 and f=0.6Hz in Figure 7. They indicate two major features of the amplification sensitivity:

1) At 0.4Hz, only the thickest part of the deposit is excited. The resonance magnitude grows up when the incidence angle increases until 80 degrees, above which the magnitude falls down again. Conversely, at 0.6Hz corresponding to the resonance frequency of the edges, the magnitude of the southern part of the deposit (left handside) is decreasing with incidence angle, but a very high amplification factor appears in the northern thinnest part of the deposit which is also increasing with the incidence angle as for the previous frequency.
2) As the magnitude changes, the peak locations shift from North to South.

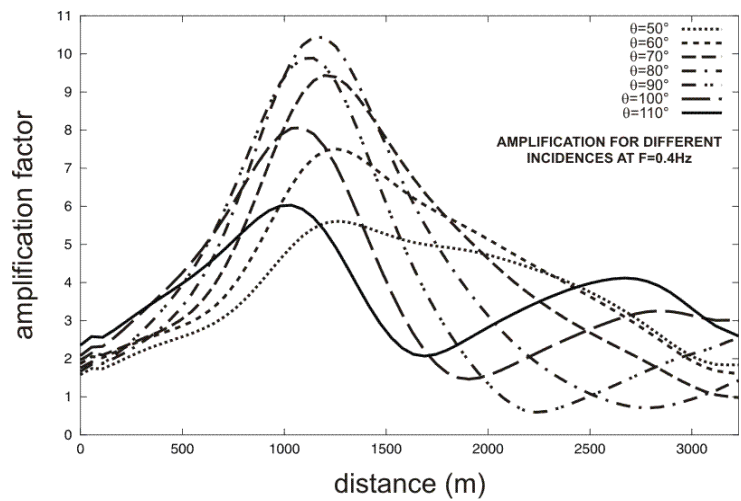

Figure 6: Amplification factor along the Tunis basin for different incidences at 0.4Hz

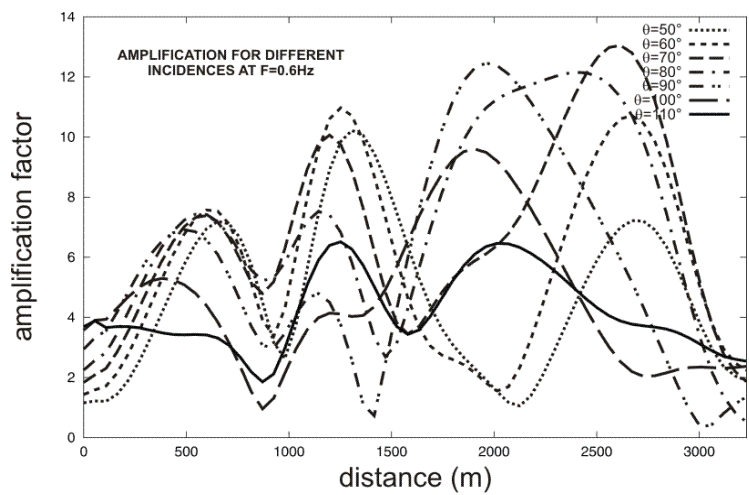

Figure 7: Amplification factor along the Tunis basin for different incidences at 0.6Hz

It can be noticed that these results are opposite to that obtained for a shallow deposit in Nice (Semblat et al, 2000a) where the main amplification factor is increasing with decreasing incidence angle at the resonance frequency. This result is mainly due to the fact that the shallow alluvial deposit in Nice can locally be considered one-dimensional. In the case of Tunis, both basin edge and focusing effects due to the irregular geometry (Bouden-Romdhane and Mechler, 2002) are certainly affected by incidence angle and then greatly modify the spatial variability of site effects.

## 5    Comparison between numerical and experimental results

In Figure 8, we compare the motion amplification at BOUR station with respect to frequency for a vertically incident SH wave to the HVSR curve and an average SSR curve taken from eight earthquakes (Bouden-Romdhane and Mechler, 2002). It must be noticed that for the SSR data, only data for which the signal-to-noise ratio is larger than three are considered, which explains a missing part in the signal (dashed line) for high frequencies.

In spite of the simplifications of the model, it is quite satisfactory to see that the main peaks observed in the experiments are recovered by the computations in terms of both magnitude and frequency, especially below 2Hz. The maximum amplifications correspond to sharp peaks at 0.4, 1, 1.25, 1.4, 1.6 and 2Hz. However, above 2Hz, a shift of the numerical peaks from the experimental ones can be noticed. It is worth noting that the amplification at the main resonance frequency at 0.4Hz is overestimated by the numerical solution compared to SSR, reaching 6 instead of 4 (Figure 8).

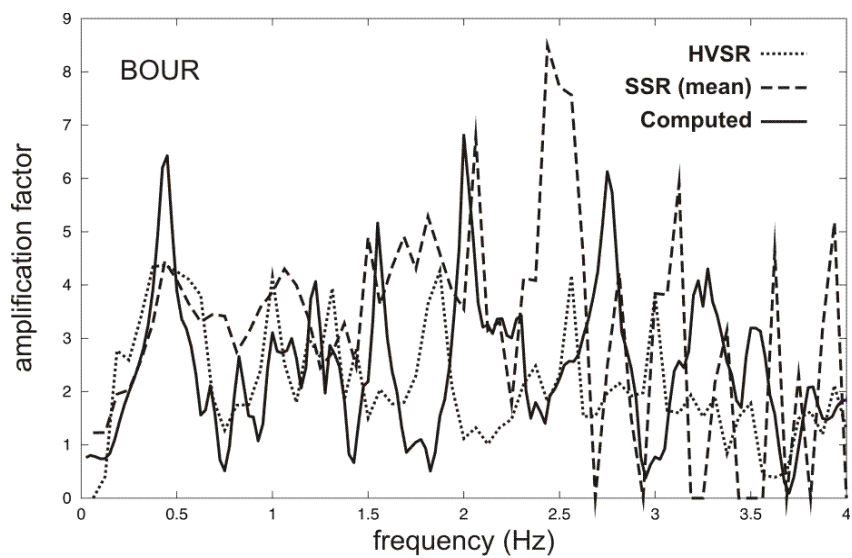

Figure 8: Amplification at BOUR station: comparison between the experimental HVSR (dotted), average SSR from 8 earthquakes (dashed) and the numerical solution (solid).

As it will be shown later, when a 5% damping ratio is considered in order to take into account the anelastic reponse of the soil, the main resonance amplification is reduced to 5. From the peaks identified in Figure 8 for the BOUR station, the amplification is now analyzed in the whole basin at these five frequencies (Figure 9). These isovalues show the different locations of amplification nodes nearby the surface of the alluvial deposit. As the frequency increases, the number of amplification nodes becomes larger and they reach deeper locations inside the deposit (Figures 9c to 9f). The strong sensitivity of the location of maximum amplification along the surface may explain the shift of the peaks for higher frequencies observed in Figure 8. In Figure 10, we replace the local amplification computed at BOUR by the overall maximum amplification along the surface as suggested by Semblat in the case of Nice for comparison with average amplification curves from experiments (Semblat et al, 2000a). The meaning of such a representation is to avoid the sensitivity on the location of maximum amplification areas. As a matter of fact, Semblat showed that these overall maximum results are consistent with several experiments. As we may observe in Figure 10: when considering the overall amplification, the shift in frequency between the numerical and experimental results disappears.

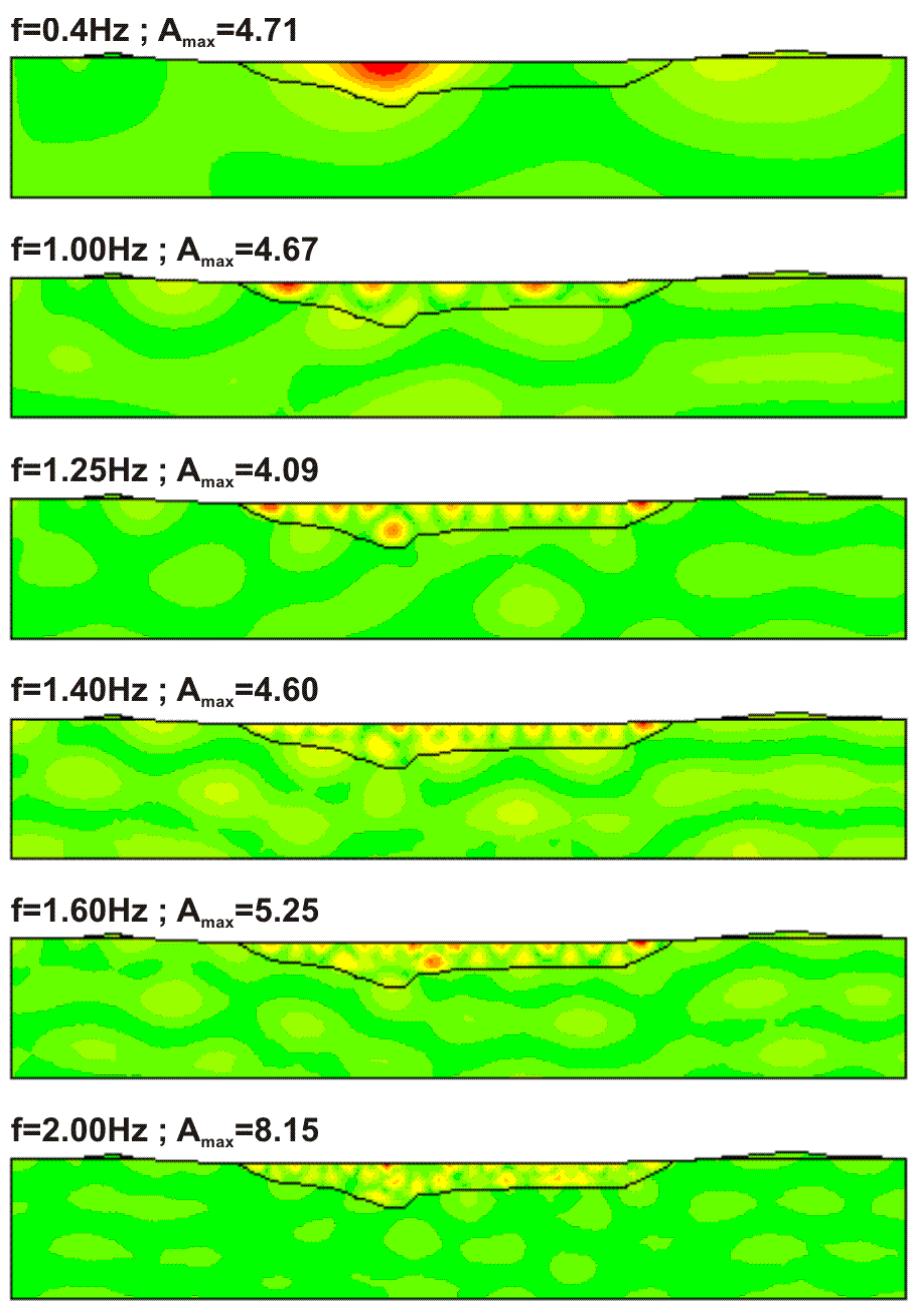

Figure 9 : Amplification factor in the Tunis basin estimated from BEM simulations at various frequencies.

Furthermore, it is worth noticing that the comparison between the numerical results and experimental ones is even better for the fundamental mode when a 5% damping is considered (Figure 10). Since a simple attenuation model was used (Semblat et al, 2000a), the higher frequency amplification is underestimated and a NCQ type model should be considered (Semblat and Pecker, 2009). All these numerical results emphasize the major role of the whole alluvial deposit on the amplification of surface motion, while the thinnest surficial part disregarded here may play a significant role only for higher frequencies.

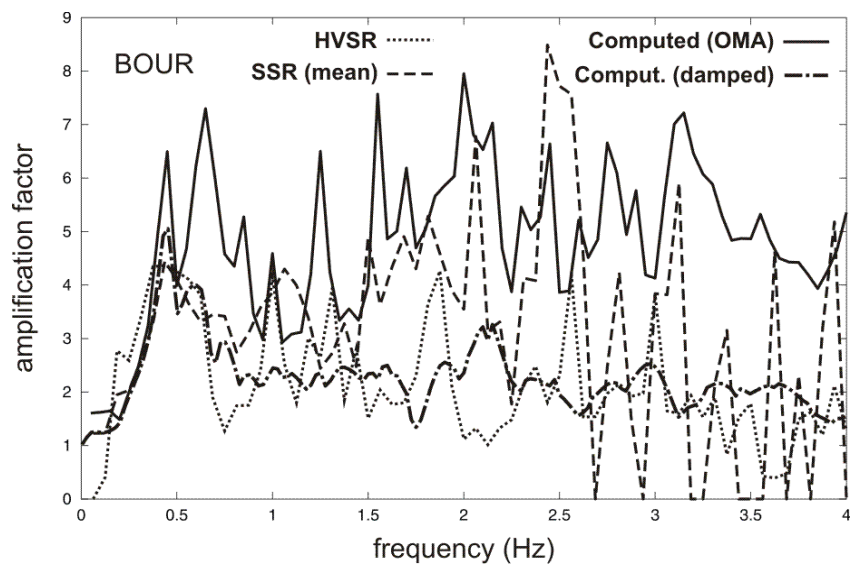

Figure 10: Measured amplification and computed overall amplification at BOUR station: comparison between the experimental HVSR (dotted), average SSR from 8 earthquakes (dashed) and the numerical solution in terms of OMA, overall maximum amplification, (solid: undamped case, dotted-dashed: damped case).

## 6   Conclusions

In the densely built area of Tunis city, site effects lead to a significant local amplification of seismic ground motion. Two experimental surveys show that the seismic motion is strongly amplified by a factor of 5 at particular frequencies below 4Hz.

Numerical simulations based on the boundary element method allow a precise description of the site as well as an accurate analysis of seismic wave propagation within the alluvial basin. Amplification levels and corresponding frequencies are of the same order as the experimental values. As also shown by experiments, the maximum amplification estimated numerically is located in the thickest part of the alluvial layer at lower frequencies and spreads to the thinnest edges for higher frequencies. Site effects quantified numerically are also sensitive to incidence angles. They can change the maximum amplification factor, the frequency of maximum amplification and the corresponding location.

The agreement between numerical and experimental results is very good especially when the overall maximum amplification is computed from the simulations. More complex damping properties also have to be included to retrieve the amplification level at the fundamental frequency of the deposit and above.

Further works should take into account several important physical parameters such as damping, wavetypes, surficial heterogeneities in the deposit, etc. However, additional experimental data would be necessary.